# One-by-one trap activation in Silicon nanowire transistors

N. Clément[1], K. Nishiguchi[2], A. Fujiwara[2], & D. Vuillaume[1]

(1) Institute of Electronics, Microelectronics and Nanotechnology, CNRS, University of Lille, Avenue Poincaré, 59652, Villeneuve d'Ascq France

(2) NTT Basic Research Laboratories, 3-1, Morinosato Wakamiyia, Atsugi-shi, 243-0198 Japan

Flicker or 1/f noise in metal-oxide-semiconductor field-effect transistors (MOSFETs) has been identified as the main source of noise at low frequency. It often originates from an ensemble of a huge number of charges trapping and detrapping. However, a deviation from the well-known model of 1/f noise is observed for nanoscale MOSFETs and a new model is required. Here, we report the observation of one-by-one trap activation controlled by the gate voltage in a nanowire MOSFET and we propose a new low-frequency-noise theory for nanoscale FETs. We demonstrate that the Coulomb repulsion between electronically charged trap sites avoids the activation of several traps simultaneously. This effect induces a noise reduction by more than one order of magnitude. It decreases when increasing the electron density in the channel due to the **electrical screening of traps. These findings are technologically useful for any FETs with a short and narrow channel.**

In electronics, noise refers to unwanted or parasites random signals overlying the useful signals. For most electronics applications such as amplifiers, memories or digital processing, metal-oxide-semiconductor field-effect transistors (MOSFETs) are the basic constituent of circuits. However, whereas scaling is required for high level of integration and increase of working speed, for instance, low frequency noise is progressively becoming a serious issue for continuous devices scaling[1]. Power spectrum current noise in MOSFETs at low frequency follows the 1/f law meaning that noise spectrum is inversely proportional to frequency f on a logarithm scale. The 1/f noise is generally interpreted as the superposition of random events of charge trapping and detrapping to defects randomly distributed in the gate oxide (e.g. $SiO_2$) near the semiconductor channel (e.g. Si) [2,3] (Fig.1a). In shrinked MOSFETs, the number of electrically active defects is reduced, and the low-frequency-noise begins to deviate from the 1/f characteristics[3-6]. Finally, in sub-micron MOSFETs (e.g. < 100 nm by 100 nm), only a few traps exist in the device, and we observe discrete switching in the drain current between two (or more) levels under constant bias conditions[2-10] (Fig.1b). These latter fluctuations, known as Random Telegraph Signal (RTS), give a Lorentzian distribution in the power spectrum current noise. In other words, 1/f noise, resulting from an averaged ensemble of individual RTS, is no longer valid in ultra-small MOSFETs.

Therefore, the electrical properties of individual RTSs needs to be understood in order to elucidate the noise behavior in such nanoscale MOSFET.

Here, we show that one-by-one activation of RTSs at room temperature can be controlled by the gate voltage of nano-scale MOSFETs. This one-by-one activation is attributed to Coulomb repulsion between trapped electrons in neighboring defects[5,11,12]. We establish equations for low-frequency noise in such nanoscale MOSFETs and we demonstrate a drastic reduction of this low-frequency-noise. Moreover, we show that the electrical screening by the electrons in the semiconductor channel reduces this one-by-one activation.

**Results**

**Device structure**

Scanning Electron Microscope (SEM) top and schematic side views of the device are shown in Fig.1c. Larges undoped silicon-on-insulator (SOI) channels with oxide thickness $t_{ox}$ = 400 nm are locally constricted by e-beam lithography and thermally oxidized to form a 40 nm thick upper oxide (see methods). The current characteristics are determined by the constricted channel whose width W and length L after the oxidation are 15 and 50 nm, respectively [13]. The Si substrate is used as the back-gate. Such a small wire channel makes the MOSFET useful as a high-charge-sensitivity electrometer with single-electron resolution[14-16] and thus suitable for a clear observation of RTS at room temperature. Electrons can be trapped by oxide defects surrounding the SOI channel and located at a tunneling distance (e.g. < 3 nm)[2]. For a basic equivalent circuit (Fig. 1c), we consider the capacitance $C_G$ between the gate and the trap site and a tunneling capacitance $C_J$ between the channel and the trap site.

**Random Telegraph Signal amplitude in a Si nanowire transistor**

Trapping and detrapping of a single electron by a single trap close to the Si channel induce a two levels fluctuation of the drain current called RTS noise due to electrostatic effect caused by the electron. This RTS noise gives access to precise information such as the trap depth (i.e. its distance in the oxide from the Si/SiO$_2$ interface) and the gate capacitance $C_G$. In this section, we focus on the analysis of RTS noise amplitude in the nanoscale FETs and, in addition to aforementioned parameters, we also introduce an effective trap charge, q*, originating from the trapped electron. Figure.2a explains the mechanism of RTS. Let us call *I* the average value of the drain current $I_d$ and *ΔI* the amplitude of RTS signal. When an electron is trapped to a defect, the electrostatic effect induced by q* shifts current $I_d$ characteristics as a function of back-gate voltage $V_{BG}$ by *ΔV$_{FB}$*. As a result, at a constant gate voltage, this shift corresponds to a small decrease in $I_d$. When an electron is detrapped from the defect, $I_d$ returns to its initial value upon electron detrapping, thus giving rise to the two-level RTS noise. When $C_G<<C_J$[17-18], simple electrostatics leads to *ΔV$_{FB}$*=q*/$C_G$ and from the slope *gm=∂I/∂V$_{BG}$* of the I-$V_{BG}$ curve (gm is the transconductance of the transistor), we get *ΔI = g$_m$ΔV$_{FB}$*.   Thus the basic RTS equation is

$$\Delta I = gm.q*/C_G \qquad (1)$$

This equation is widely used for flat band voltage $V_{FB}$ fluctuation[17] as well as for quantifying the behavior of single-electron memories[14,18]. The effective charge $q^*$, which is used instead of the electron unit charge $q$, depends on $V_{BG}$ and will be discussed later with Fig.4c. Basically, we consider $q^*=q$ at low gate voltage, i.e. in the subthreshold region[19] ($V_{BG} < V_{FB} = 12\ V$ here), and $q^*<q$ above $V_{FB}$. Figure 2b shows two examples of RTSs measured in our Si nanowire MOSFET. The upper panel shows a basic two-level current fluctuation behavior, and the lower one shows a more sophisticated case of a three-level current fluctuation that will be discussed later (see section Coulomb repulsion analysis). Figure 2c shows $I$-$V_{BG}$, $\Delta I$-$V_{BG}$, and $gm$-$V_{BG}$ characteristics. We distinguish the RTS contributions of 5 traps in $\Delta I$-$V_{BG}$ curve from a detailed analysis of the time dynamics of the RTS signal (see next section). Then, from data in Fig.2c (limited below $V_{FB}$=12 V to assume $q^*=q$) and from Eq.1, we can estimate $C_G = 0.91\pm0.18$ aF, which is a realistic value from the viewpoint of devices geometry. This proves the validity of the simple eq.1 related to RTS amplitude. Such a clear RTS amplitude dependence with the transconductance gm for several traps is obtained because the oxide thickness between the channel and gate is much larger than the trap depth (distance from the Si nanowire, see Fig.1c), i.e., $C_G$ much smaller than $C_J$, and because the devices dimensions (in particular the width) are much smaller than the Debye screening length (here ~ 110 nm at room temperature and for a Si nanowire doped at $10^{15}$ cm$^{-3}$), otherwise, the trapped charges in the defects would lead to a more complicated RTS amplitude analyzis. As a consequence, such RTS measurement in a nanoscale

MOSFET is a highly effective metrology tool for evaluating gate capacitances in the sub attofarad range, which is very difficult with other techniques.

**Traps occupancy probabilities**

To identify the RTS signal of each trap site, we record the time duration of the high and low currents for a large number of such RTS fluctuations, as shown in Fig.2b. According to standard statistical analysis[2], we deduce the average electron capture time ($\tau_c$) and emission time ($\tau_e$), respectively. Figure 3a shows the dependence of $\tau_c$ and $\tau_e$ as a function of $V_{BG}$. We can identify four sets of $\tau_e$ and $\tau_c$ at different values of $V_{BG}$, corresponding to four traps and, more interestingly, that the four RTSs become active in turn when increasing $V_{BG}$. Hereafter, we refer to those sites as trap 1 to 4 as shown in Figs. 2c and 3a. For trap i (i = 1 to 4), the $V_{BG}$ dependence of the probability that a trap is occupied by an electron $g_i = \tau_e/(\tau_e + \tau_c)$, also shows a clear one-by-one activation of RTSs. The behavior of such one-by-one activation of RTSs seems to be unnatural in the well-known RTS theory because it means that trap sites have well align energy levels although expected to be randomly distributed in space and energy. We will show below that Coulomb repulsion between electrons trapped by a defect can satisfactorily explain this behavior.

**Coulomb repulsion analysis**

For the analysis of the Coulomb repulsion, let us return to the three-levels RTS shown in Fig. 2b that is observed in the bias range 20.5 V < $V_{BG}$ < 23.5 V. The three-levels RTS implies two different traps active simultaneously with same *ΔI*. It is attributed to trap 4 and another trap (trap5). Upper (U), middle (M) and lower (L) levels means that no trap, only one trap and both traps are filled by electrons, respectively. In this bias range, the histograms of U, M and L levels in $I_d$ follow Gaussian distribution with a relative amplitude that depends on $V_{BG}$. Figure 3c shows examples at $V_{BG}$ = 20.5, 22, and 23.5 V. Since the amplitudes of the peaks depend on $V_{BG}$, normalization of each amplitude allows evaluation of probabilities $P_U$, $P_M$, and $P_L$ of U, M, and L levels, respectively, in three-level RTSs (Fig. 3d). With g4 and g5 the occupancy probabilities of traps 4 and 5, they can be given simply by $P_U$=*(1-g4)(1-g5)*, $P_M$=*g4(1-g5)+g5(1-g4)*, and $P_L$=*g4g5* for the trap conditions illustrated in the inset of Fig. 3d. To evaluate $P_U$, $P_M$ and $P_L$, we first use the usual g partition function for traps 4 and 5 (g4 and g5):

$$g_i = \frac{1}{1+e^{\frac{q}{kT}\left(E_{Ti}-\frac{C_G}{C_{Ji}}V_{BG}\right)}} \qquad (2)$$

where $E_{Ti}$ is the difference between trap potential energy (i=1 to 5) and Fermi energy at $V_{BG}$ = 0 V, $C_{Ji}$ is $C_J$ of trap i, $C_G$ the back-gate capacitance (see Fig.1d), k the Boltzman constant, T the temperature, q the electron charge and $V_{BG}$ the back-gate voltage. However, eq.2 leads to poor fits (see supplementary information, section 1). To obtain satisfactory fits for $P_U$, $P_M$ and $P_L$, we use

$$g'_i = \frac{1}{1+e^{\frac{q}{kT}\left(E_{Ti}+g'_j\varphi_{ij}-\frac{C_G}{C_{Ji}}V_{BG}\right)}} \qquad (3)$$

by introducing an additional term $g'_j \varphi_{ij}$ (Eq.1). The $g'_j \varphi_{ij}$ corresponds to the Coulomb repulsion potential between trap i and trap j, weighted by occupancy probability $g'_j$ of the interacting trap j. Using Eq. 3 instead of Eq.2 for the probabilities $P_U$, $P_M$ and $P_L$ allows us obtaining a reasonable fit as shown in Fig.3d. From this analysis, we extract traps 4 and 5 occupancy probabilities $g'_4$ and $g'_5$ reported in Fig.3b.

The qualitative meaning of this modified equation (Eq. 3) can be simply illustrated with a band energy diagram (Fig. 4a) for traps 4 and 5 causing the three-level RTS. The key point is a competing effect, quantified by $h_i$, of Coulomb repulsion $g_j \varphi_{ij}$ and $V_{BG}$-induced potential drop $C_G V_{BG}/C_{Ji}$ of the trap energy level, that can be given by eq.4.

$$h_j = \frac{d}{dV_{BG}} \left[ \sum_{j \neq i} g'_i \varphi_{ij} - \frac{C_G}{C_{Jj}} V_{BG} \right] . \qquad (4)$$

In the range of $V_{BG}$ between 18 to 23.5 V, the energy level of trap 4 aligns close to that of trap 5, which leads to three-level RTSs. However, since trap 5 is located deeper from the channel than trap 4 as shown in table 1 (see methods for the determination of trap depth), removing an electron at trap 5 is harder than removing one at trap 4. More importantly, the Coulomb repulsion is larger than the potential drop of trap sites caused by the applied bias $V_{BGs}$, that is, $h_5$ is positive. Therefore, in this $V_{BG}$ range (18-23.5 V), we can now describe the coupled behavior of traps 4 and 5 (Fig.3b). As trap 4's occupancy probability increases with $V_{BG}$, trap 5's energy level is pushed up (kink in g'5). Since it is not enough to get complete blockade, trap 5's occupancy probability increases and pushes up trap 4's energy level (decrease of g'4). When trap 5 is almost always filled, trap 4's occupancy probability increases again (increase of g'4). The estimated Coulomb repulsion between trap 4 and 5, $\varphi_{45}$ = 110 meV (see table 1).

This Coulomb effect also explains the one-by-one activation of RTSs as shown in Fig.3b, i.e. the fact that there is no overlap of the occupancy functions of traps 1, 2 and 3. The data are well separated along the $V_{BG}$ axis and Eq.3 reproduces well this behavior. It means that the shaded areas in Fig.3b correspond to high Coulomb repulsion as indicated by the positive values of $h_j$ (Fig.4b) for traps 1, 2 and 3. The more complicated curve for $V_{BG}$ > 18 V corresponds to the interacting behavior between traps 4 and 5 as discussed above. As a consequence, it could be concluded that energy levels of traps when empty are close to each other which is natural and more feasible given a similar chemical origin. These considerations based on Eqs. 3 and 4 can explain one-by-one activation of RTSs shown in Fig. 3b, i.e., no overlap, (shaded areas in Fig. 3b), between each $g'_i s$ at positive $h_j$ in particular $V_{BG}$ regions of Fig. 4b.

This idea can also be explained by an analysis on $\varphi_{ij}$. From the data shown in Fig. 2c, using Eq.1 and assuming $C_G$ is not depending on $V_{BG}$[20] ($C_G$ = 0.91 aF± 0.18 aF, see above), we can calculate the effective charge q* for each trap (table 1). The value of q* decreases with increasing $V_{BG}$ after the channel inversion. It is interesting to note that the same behavior is obtained for the normalized values of $\varphi_{ij}$, i.e. $\varphi_{ij}/\varphi_{12}$. Figure 4c shows this comparison. These features can be explained by considering traps image charge[21] and an electrical screening effect originating from the reduction of charges in the inversion layer of the channel, that allows $\varphi_{ij}$ and q* to be given by (see methods)

$$\varphi_{ij} = \frac{q*}{4\pi\varepsilon_1 r_{ij}} = \frac{q}{4\pi\varepsilon_1 r_{ij}} \left[ \frac{1}{1 + \frac{qt_{acc}}{4kT\varepsilon_1} \frac{C_G}{WL}(V_{BG} - V_{FB})} \right] \quad (5)$$

, where $\varepsilon_1$ is the dielectric constant of $SiO_2$, $r_{ij}$ the distance between two trap sites and $t_{acc}$ the thickness of the inversion layer of the channel. Figure 4c shows the good agreement between the behavior of $\varphi_{ij}$ and Eq.5 with $t_{acc}$ and $r_{ij}$ equal to 0.4 and 2 nm, respectively. This result means that all traps are located within a few nanometers of each other.

**Derivation of power spectrum noise equations**

More interestingly and importantly, the above detailed analysis of q* and $\varphi_{ij}$ based on Coulomb repulsion provides us a better understanding of low frequency noise in nanoscale MOSFETs[19] and especially its deviation from the well known 1/f noise. We address this noise issue that can have usefull implications for design and simulation of nanoscale MOSFETs. In large devices, with the assumption that 1/f noise is composed of an ensemble of a large number of RTSs originating from traps randomly distributed in space and potential level, the *1/f* power spectrum current noise $S_{I1}$ is given by[17,19]

$$\frac{S_{I1}}{gm^2} = N\left(\frac{q*}{C_G}\right)^2 \frac{1}{f} \quad (6)$$

where N is the number of active traps. This equation predicts an increase of noise when decreasing the device area S, because $C_G$ and N (at a given trap density) scale with S. We measured the low-frequency noise at different $V_{BG}$ (see Methods). A typical curve measured at $V_{BG}$ = 16 V is shown in Fig.5a. The measured noise deviates from a strict 1/f noise and it is composed of a 1/f background noise superimposed by a Lorentzian shape related to the RTS noise generated by trap 3 for this peculiar bias $V_{BG}$ = 16 V. An ultimate lower limit of Eq.6 calculated for N=1 (one trap, albeit strictly speaking not valid for Eq.6) with the gm, q* and $C_G$ values for the same $V_{BG}$ (trap 3) clearly overestimates the noise amplitude compared to the experimental data (Fig. 5a). When only a few traps are present in nano-scale devices, a better approach is to use the Machlup derivation[2,6,22] of the Lorentzian equation from RTS. Thus, we can express the low frequency power spectrum $S_{I2i}$ for trap i by (see methods).

$$\frac{S_{I2i}}{gm^2} = \frac{4\,g_i\,(1-g'_i)^2\,\tau_{ei}(q*/C_G)^2}{[1+(2\pi\,(1-g'_i)\,\tau_{ei}\,f)^2]} \tag{7}$$

We calculated this quantity at 10 Hz for each traps, i=1 to 5, as a function of $V_{BG}$ using traps parameters (q*, $C_G$) and $g'_i$ functions given in Table 1 and Fig.3b. Each curve is in good agreement with the experimental data. Here $\tau_{ei}$ is considered constant with average value $\tau$ listed in Table 1. The results are shown by the bell-shaped curves in Fig.5b and are compared with the experimental data. Each calculated curve is in good agreement with the experimental data. Note that traps 2 and 3 contribute to the noise spectra at 10 Hz for the same range of $V_{BG}$ but with a negligible contribution for trap 2. This result is due to the fact that trap 2 has a higher time constant than trap 3 (see Fig.3a). As a consequence, the experimental data around $V_{BG}$ = 10 V comes from the 1/f background noise also observed in Fig.5a. The discussion of the physical origin of this noise is out of the scope of this paper. We suggest elsewhere that it should be due to the dipolar polarization noise in the oxide[23]. For the sake of device simulation, Eq.7 is not very practical since it requires a detailed knowledge of the physical parameters of all defects involved in the device. A simplified expression of the maximum of eq.7 can be derived (see Methods),

$$\frac{S_{I\max}}{gm^2} = 0.08 \frac{(q*/Cg)^2}{f} \tag{8}$$

Eq.8 can be used to estimate an upper limit of noise as shown in Figs.5a and 5b. Especially, a $g'_i$ value of 0.5 in Eq.7 gives a good estimation of the noise $SI_{max}$ at the corner frequency of the Lorentzian spectrum as shown in Fig. 5a. This means that when just a few traps are active in nanoscale MOSFETs, the classical equations for 1/f noise (eq.6) should be still used in device simulation if compensated by a correction factor of about 0.08 as an upper approximation (eq.8).

### Discussion

Among the ten measured samples, two did not have any RTS, six had a single RTS and two had many traps (RTSs) with one-by-one trap activation, one of which has been presented here. The second one is shown in supplementary information, section II. For a high quality thermal oxide, a typical density of oxide traps is about $10^{10}$ cm$^{-2}$ in an energy window of kT. These traps are mainly related to dangling bonds in $SiO_2$ and at the $Si/SiO_2$ interface[24]. Our device has a surface area of about 15 nm x 50 nm and an energy window $C_G \Delta V_{BG}/C_{Javg} \approx 0.7$ eV, where $C_{Javg}$ is the average of $C_{Ji}$ in table 1 and $\Delta V_{BG}$ is the window of back-gate voltages. Therefore, statistical number of traps in our device can be estimated to 2.1, which is not so far from the experimental results from the statistical viewpoint.

One-by-one activation of RTSs demonstrated here and the corresponding noise reduction should be also relevant for any other NW-based devices, such as carbon nanotubes and other bottom-up compound semiconductor wires. However, the noise reported is still high [25,26], compared to state-of-the-art Si MOSFETs. This is because the nanotube in these devices is about a micrometer long, and thus there are a lot of trap sites that have no interaction (i.e., Coulomb repulsion) between them. Therefore, for the lasting benefit of noise reduction, nanowire devices should have a few ten nanometer length, in which Coulomb repulsion between charges located at nearby trapping sites is effective.

K. R. Farmer et al. have reported the correlation between two RTS events[27]. However, they didn't give any detailed analysis for evaluating the RTS amplitude $\Delta I$, the trap occupancy probabilities, the location of trap sites, the Coulomb term $\varphi_{ij}$, the channel carrier screening effects, nor the influence on power spectrum noise. We report here such an analysis, because the extremely small gate capacitor of the SOI-based MOSFETs enhances the Coulomb repulsion. Our detailed analysis allows a step-by-step evaluation of the above key parameters of each traps active in the nanoscale FET and we propose a new model for the low frequency noise of this nanoscale device suitable for device simulator. Therefore, this nanoscale-MOSFET with a short channel and small gate capacitor can be used as a metrological tool for the analysis of capacitances and low-frequency noise. These approaches can be extended as well to devices composed of carbon nanotubes, graphene nanoribbons, and any other state-of-the-art nanoscale structures.

**METHODS.**

**Device fabrication**

The nanoscale MOSFETs were fabricated on a silicon-on-insulator (SOI) wafer. First, a narrow constriction sandwiched between two wider (400-nm-wide) channels was patterned on the 30-nm-thick top silicon layer (p-type, boron concentration of $10^{15}$ cm$^{-3}$). The length and width of the constriction channel was 30 and 60 nm, respectively (Fig.1c). The patterning was followed by thermal oxidation at 1000 °C to form a 40-nm-thick $SiO_2$ layer around the channel. This oxidation process reduced the size of the constriction to about 15 nm, giving a final channel dimension of 15 x 50 nm. Then, we implanted phosphorous ions outside the constriction, five micrometer away from it using a resist mask, to form highly doped source and drain regions. Finally, aluminum electrodes were evaporated on these source and drain regions.

**Electrical measurements**

Electrical measurements were performed at room temperature in a glove-box with a controlled $N_2$ atmosphere (< 1 ppm of $O_2$ and $H_2O$). Drain voltage $V_D$ (usually 50 mV) and back gate voltage (< 8V) were applied with an ultralow-noise DC power supply (Shibasoku PA15A1 when $V_{BG}$ < 8 V or Yokogawa 7651 when $V_{BG}$ > 8 V). The source current was amplified with a DL 1211 current preamplifier supplied with batteries. RTS data and noise spectra were acquired with an Agilent 35670 dynamic signal analyzer.

**Determination of oxide trap depths**

The trap depth are estimated by fitting with eq.3 the experimental trap occupancy. $C_{Ji}$s obtained for each trap are reported in Table 1. In a parallel plate configuration, $C_{Gi} \approx yt_i/t_{ox}$, where $yt_i$ is the trap depth and $t_{ox}$ is the gate oxide thickness. $t_{ox}$ = 400 nm is larger than the width W and length L of the nanowire and we cannot neglect border effects. This induces a correction factor of 7.5. Therefore, $yt_i \approx C_G.t_{ox}/ (C_{Ji}.7.5)$ as reported in Table.1.

**Theoretical derivation of Eq.5**

If $V_{BG}$>$V_{FB}$, an accumulation layer appears in the Si channel at the $SiO_2$

interface. This affects the dielectric properties of the SiNW. An effective dielectric constant for Si is introduced to consider effects of screening by electrons in the channel. Considering the accumulation charge $Q_{acc}$ related to capacitance $C_{acc}$, surface potential $\psi_s$, the Debye screening length $L_d$ and accumulation layer thickness $t_{acc}$, we write

$$Q_{acc} = \frac{\varepsilon_0 \varepsilon_2 kT}{qLd} WL e^{\frac{e\psi_s}{2kT}} = C_G(V_{BG} - V_{FB}) \tag{9}$$

$$C_{acc} = \frac{\partial Q_{acc}}{\partial \psi_s} = \frac{q}{2kT} C_G(V_{BG} - V_{FB}) = \frac{\varepsilon_0(\varepsilon'_2 - \varepsilon_2)}{t_{acc}} WL \tag{10}$$

where W and L are the width and length of the NW, $C_G$ the back-gate capacitance (see Fig.1c), k the Boltzmann constant, T the temperature, q the electron charge, $V_{BG}$ the back-gate voltage, $V_{FB}$ the flat-band voltage, $\varepsilon_0$ the vacuum permittivity, $\varepsilon_2$ the Si relative dielectric constant and $\varepsilon'_2$ the effective Si dielectric constant.

From Eq. 10, we get

$$\varepsilon'_2 = \varepsilon_2 + \frac{q}{2kT\varepsilon_0} \frac{t_{acc}}{WL} C_G(V_{BG} - V_{FB}) \tag{11}$$

$$\varphi_{ij} = \frac{q}{4\pi\varepsilon_0 \varepsilon_1 r_{ij}} + \frac{(\varepsilon_1 - \varepsilon_2)}{(\varepsilon_1 + \varepsilon_2)} \frac{q}{4\pi\varepsilon_0 \varepsilon_1 r'_{ij}} \tag{12}$$

with $\varphi_{ij}$ the electric potential at a distance $r_{ij}$ to the trap and $r'_{ij}$ to its image charge[21]. If we consider $r_{ij} \approx r'_{ij}$ (in other words, the distance between traps large compared to trap depth in oxide), we have

$$\varphi_{ij} = \frac{q}{4\pi\varepsilon_0 \frac{(\varepsilon_1 + \varepsilon'_2)}{2} r_{ij}} \quad (13)$$

Combining Eqs. 11 and 13

$$\varphi_{ij} = \frac{q}{4\pi\varepsilon_0 \varepsilon_1 r_{ij}} \left(\frac{2\varepsilon_1}{\varepsilon_1 + \varepsilon_2}\right) \left[\frac{1}{1 + \frac{qt_{acc}}{2kT\varepsilon_0(\varepsilon_1 + \varepsilon_2)}\frac{C_G}{WL}(V_{BG} - V_{FB})}\right] \quad (14)$$

Then the effective trapped charge q* is given by

$$\varphi_{ij} = \frac{q^*}{4\pi\varepsilon_0 \varepsilon_1 r_{ij}} \quad (15)$$

$$\frac{q^*}{q} = \frac{2\varepsilon_1}{(\varepsilon_1 + \varepsilon_2)}\frac{\varphi_{ij}}{\varphi_{12}} = \frac{2\varepsilon_1}{(\varepsilon_1 + \varepsilon_2)} \left[\frac{1}{1 + \frac{qt_{acc}}{2kT\varepsilon_0(\varepsilon_1 + \varepsilon_2)}\frac{C_G}{WL}(V_{BG} - V_{FB})}\right] \quad (16)$$

In this work, the Debye screening length of the undoped silicon is very large (>> 100 nm) compared to Si thickness (15 nm thick Silicon on Insulator), so we consider $\varepsilon_2 \approx \varepsilon_1$ and Eq.14 is reduced to Eq.5 used in the text.

**Theoretical derivation of Eqs. 7 and 8**

Starting from the Machlup derivation[22] of power spectrum noise:

$$S_I = \frac{4(\Delta I)^2}{(\tau_c + \tau_e)\left[\left(\frac{1}{\tau_c} + \frac{1}{\tau_e}\right)^2 + (2\pi f)^2\right]} \quad (17)$$

with ΔI the RTS amplitude, $\tau_e$ and $\tau_c$ the trap emission and capture times, f the frequency. Considering g = $\tau_e$ / ($\tau_c+\tau_e$) and ΔI/gm = q*/$C_G$ (see Eqs. 1 and 2 in the main text), we get

$$\frac{S_I}{gm^2} = \frac{4g(1-g)^2 \tau_e (q^*/C_G)^2}{[1+(2\pi(1-g)\tau_e f)^2]} \quad (18)$$

At the corner frequency of the Lorentzian distribution: $2\pi f(1-g)\tau_e = 1$ and considering g = ½, eq. 18 becomes

$$\frac{S_I}{gm^2} = \frac{(0.5)^2 (q^*/Cg)^2}{\pi f} = \frac{0.08 (q^*/Cg)^2}{f} \quad (19)$$

**ACKNOWLEDGEMENTS.**

We sincerely thank C. Delerue from IEMN for theoretical discussions on Coulomb repulsion and R. Leturcq, S. Pleutin, F. Alibart of IEMN, and I. Mahboob of NTT for careful reading of the manuscript.

**AUTHOR CONTRIBUTIONS.**

K. N. fabricated the devices. N. C. performed the electrical measurements. N. C. and K. N. analyzed the data. N. C., K. N, A. F and D. V. discussed the results and wrote the paper.

**AUTHOR INFORMATION.**

The authors declare no competing financial interests. Correspondence and requests for materials should be addressed to N. C. (nicolas.clement@iemn.univ-lille1.fr) or K. N. (knishi50@aecl.ntt.co.jp).

Supplementary Information accompanies this paper at www.nature.com/naturecommunication

CAPTIONS.

**Fig.1 General description of trapping/detrapping noise in transistors.**

Schematic view of a conventional (a) and nanowire (b) MOS transistor with examples of current fluctuations in the time domain and the corresponding power spectrum noise. In nano-MOSFETs, fluctuation of the current due to trapping-detrapping of an electron in a defect leads to discrete steps called random telegraph signal (RTS) at room temperature and a Lorentzian power spectrum noise. Due to the huge number of traps located at different depths in the oxide in conventional transistors, 1/f law is obtained as a sum of Lorentzian spectra with different corner frequencies. c Scanning Electron Microscope image of the Si nanowire and source and drain regions (top view) and schematic side view of an SOI MOSFET with a constriction with dimensions W and L of 30 and 60 nm, respectively, before thermal oxidation. The oxidation process reduces the size of the constriction W to about 15 nm (see Methods). In order to observe the RTS, we chose a MOSFET with few trap sites in the gate oxide surrounding the SOI channel (some MOSFETs show no RTS characteristic due to their small and high-quality channel). The equivalent circuit is given by two capacitors $C_G$ between the Si back-gate and the trap site and $C_J$ between the constricted channel and the trap site. Since the trap site is close to the channel, an electron is trapped there by a tunneling event through a thin oxide layer. In order to distinguish $C_J$ from $C_G$, we call $C_J$ a tunneling capacitance and draw it in a way different from conventional capacitance $C_G$ as show in the figure.

**Fig.2 Random Telegraph Signals in a Si nanowire transistors**

a. Schematic description of the impact on drain current Id of the trapping-detrapping of an electron in a defect. The two top side schematic views of the SiNW transistor show an example of defects without electrons (left) and with electron (one trap filed: right). Schematic $Id$-$V_{BG}$ curves for both cases are shown. When an electron is trapped in a defect, it induces a shift of threshold voltage (red curve). At a given $V_{BG}$, the current decreases suddenly as shown in the bottom figure. This effect is reversible. b. Typical RTSs observed in the drain current flowing through a MOSFET at $V_D$ of 50 mV and room temperature. Top: Example of two-level RTS at $V_{BG}$ of 4 V. Capture and emission time corresponds to residency time in the upper and lower levels, respectively. Bottom: Three-level RTS at $V_{BG}$ of 22.5 V. U, M, L corresponds to the upper, middle, and lower levels in the RTS. In both figures, dotted lines are guides for the eyes for recognition of the two- or three levels in Id. c $Id$-$V_{BG}$ and $gm$-$V_{BG}$ curves are plotted with large and small black circles, respectively. The RTS amplitude $ΔI$-$V_{BG}$ curves (colored circles) for four different traps show almost the same behavior as $g_m$-$V_{BG}$. Each RTS coming from its effective trap site is identified as trap 1 to trap 5 from analysis of the trapping/detrapping dynamics as shown in Fig. 3. Since traps 4 and 5, which give the same $ΔI$, are active in the same bias range between 20 and 23.5 V (three-level RTS), the number of points is doubled in this range.

**Fig.3 Traps occupancy probabilities**

a Average emission ($\tau_{ei}$) and capture ($\tau_{ci}$) interval for trap i (i=1 to 4) as a function of $V_{BG}$. b: Traps occupancy probabilities for 5 traps. Closed circles are experimental values derived from $\tau_c$ and $\tau_e$ following Eq.2. Solid curves for traps 1-5 are fitted to the experimental values by using Eq. 3. Shaded areas are $V_{BG}$ ranges where strong Coulomb repulsion between electrons located at traps is obtained from Eq. 4 as plotted in Fig.4b. c: Histograms of $I_d$ for $V_{BG}$ = 20.5, 22 and 23.5 V. d: Probabilities P(U), P(M), and P(L) of the upper (U), middle (M), and lower (L) levels, respectively, in three-level RTSs plotted as a function of $V_{BG}$. Probabilities are obtained by normalizing each peak amplitude. Closed circles are experimental data and solid curves are fitted to experimental results using Eqs. 3 and 4 with parameters shown in Table 1. The fit for P(M) can be decomposed into two dashed curves; one for the probability that one electron is trapped at trap 4 and the other that one electron is trapped at trap 5. The inset shows the occupancy status of traps 4 and 5 corresponding to each RTS level. Closed and open circles mean that one or no electron, respectively, is located at the trap site.

**Fig.4 Coulomb repulsion and trap effective charge**

a: Energy band diagram of traps 4 and 5. When $V_{BG}$ is applied, the potential drop in the oxide $C_G V_{BG}/C_J$ lowers the trap's energy levels from a dotted black line to a dotted red line. Coulomb repulsion caused by the electrons in trap 4 (trap 5) increases the energy level of trap 5 (trap 4) by $g_4 \varphi_{45}$ ($g_5\varphi_{45}$) as indicated by a blue line. Therefore, the final energy level of trap sites is raised to a solid black line. b: $h_i$-$V_{BG}$ characteristics given by Eq. 4. $h_i > 0$ means that there is a strong repulsion coming from other traps i.e. only one trap is active at the same time. This repulsion area is reported in traps occupancy probability in Fig.3b. c: Normalized effective charge obtained by RTS amplitude (closed circles) and normalized Coulomb repulsion potential (closed stars) with $\varphi_{12\_max}$ = 220 mV as a function of $V_{BG}$. The solid curve given by Eq. 5 fits the experimental results well (with $t_{acc}$=4 Å and $V_{FB}$=12V). Dotted lines are error margins (+/- 20%, corresponding to the accuracy of the determination of $C_G$) of the fit.

**Fig.5 Contribution of each trap to power spectrum current noise**

a: Experimental $S_I$ (closed circles) as a function of frequency f. $S_I$ shows the Lorentzian spectrum, which corresponds to the RTS originating from trap 3 superimposed to a 1/f background noise (yellow area). The orange and gray lines are given by eqs. 6 and 8, respectively. b: Experimental $S_I/gm^2$ (closed circles) at 10 Hz and comparison with different theories. The bell-shaped curves are given by Eq.7 which takes into account the Lorentzian spectrum shape for each trap (filled areas with a different color for each trap) and Eq.8 (gray curve), which is a simplified equation giving an upper limit for $S_I$. A guide for experimental 1/f background noise is plotted as a dark dotted line and yellow filling. Note that the shift in $V_{BG}$ for trap 2 is due to the high time constant for this trap i.e. its Lorentzian corner frequency is << 10 Hz, which is not the case for other traps.

|  | Trap 1 | Trap 2 | Trap 3 | Trap 4-1 | Trap 4-2 | Trap 5 |
|---|---|---|---|---|---|---|
| $q^*$ | 1 | 1 | 0.7 | 0.56 | 0.56 | 0.56 |
| $\tau$ (s) | 0.1 | 20 | 0.05 | 0.05 | 0.05 | ≈ 0.05 |
| $C_{Ji}$ (aF) | 42 | 35 | 30 | 30 | 30 | 28 |
| $y_{Ti}$ (nm) | 1.15 | 1.39 | 1.6 | 1.6 | 1.6 | 1.73 |
| $E_{Ti}$ (meV) | 100 | 320 | 475 | 600 | 710 | 630 |
| $\varphi_{ij\_max}$ (meV) | 220 | 220 | 155 | 125 | 110 | 110 |

*Table 1: Extracted parameters from RTS*

Average effective trap charge, time constant $\tau$ at $\tau_e=\tau_c$, $C_{Ji}$ used in Eq.3, trap depths $y_{Ti}$ (see methods), traps energy levels $E_{Ti}$ with regard to Si Fermi level at $V_{BG}$ = 0 V from Eq.2, and difference in energy $\varphi_{ij\_max}$ between two adjacent levels.

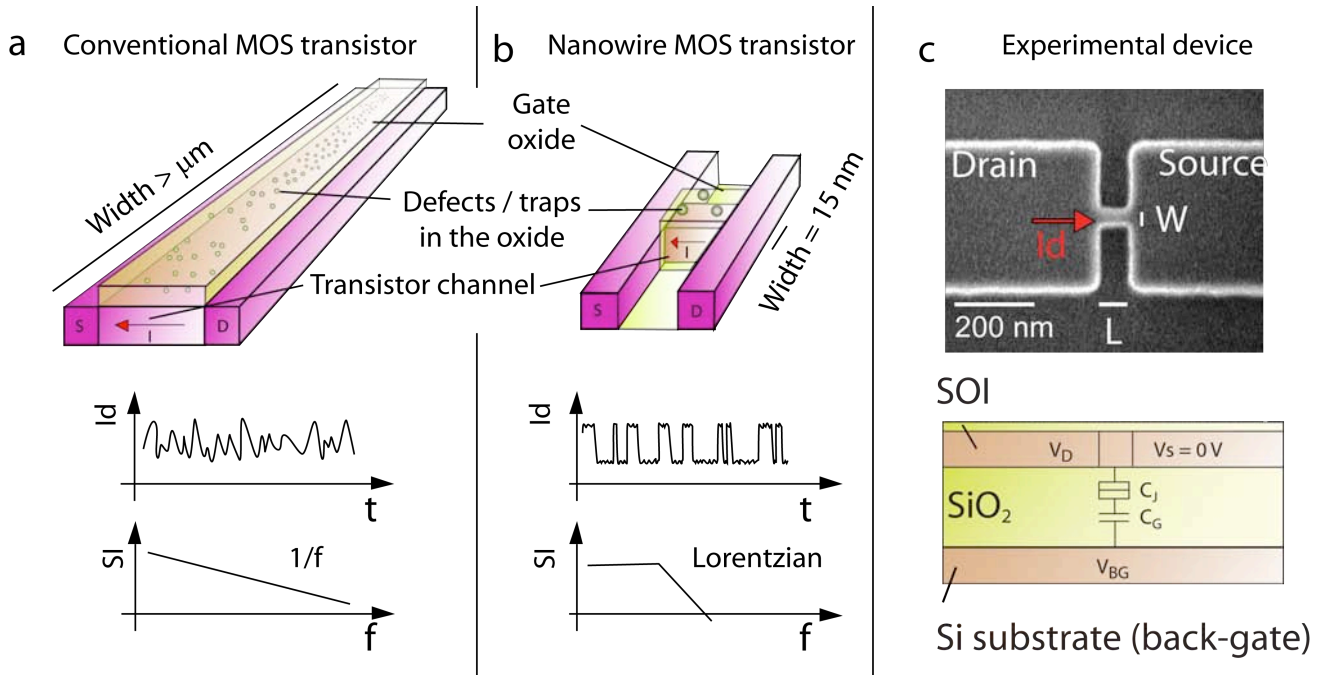

Fig.1

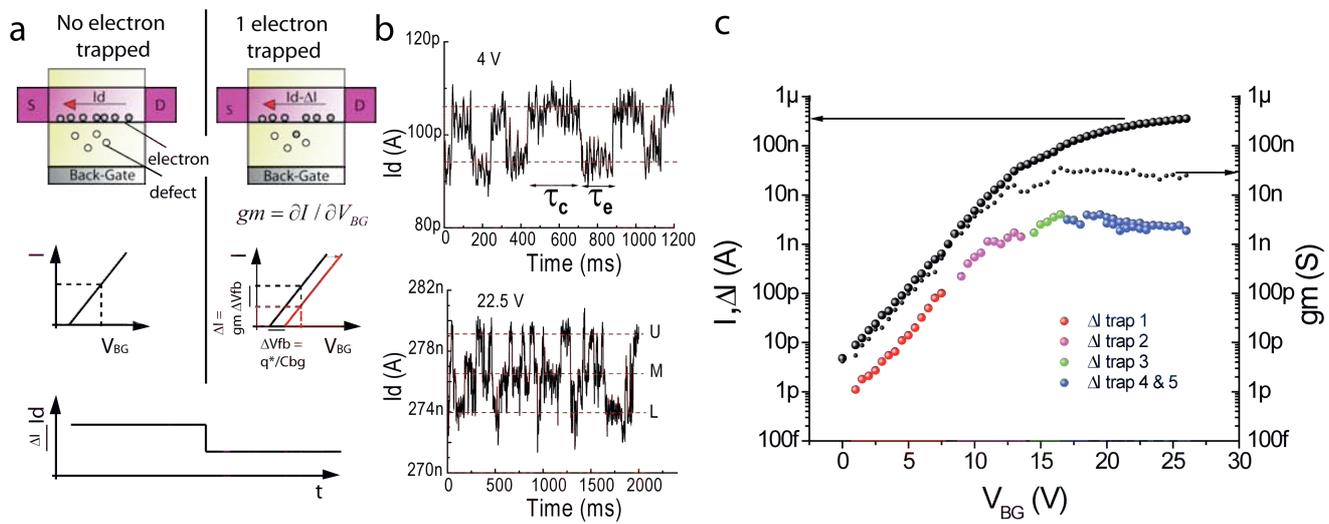

Fig.2

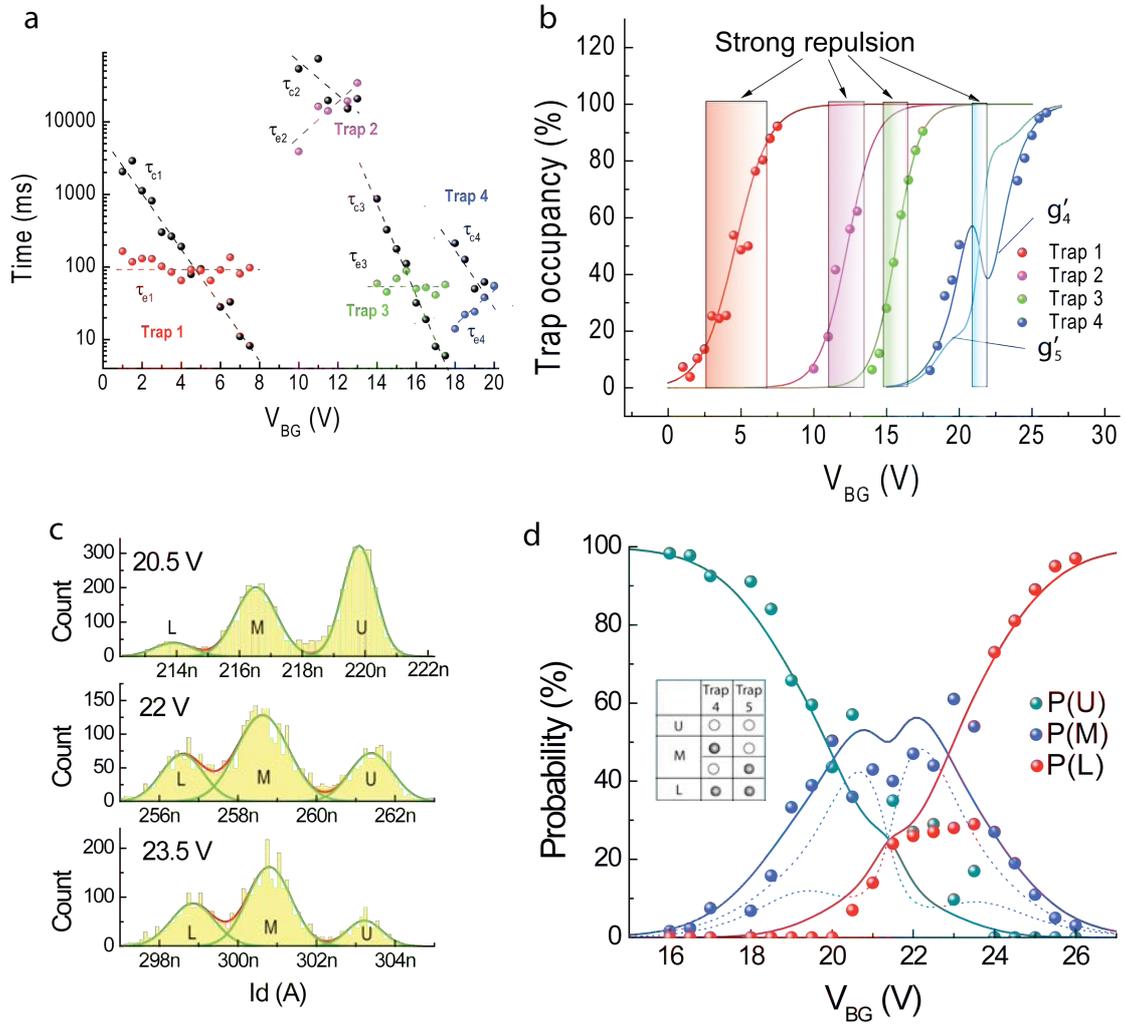

Fig.3

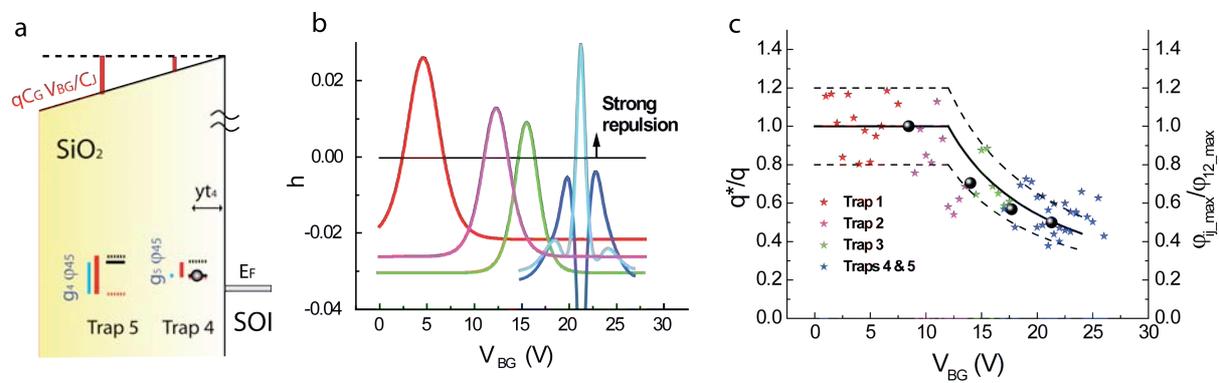

Fig.4

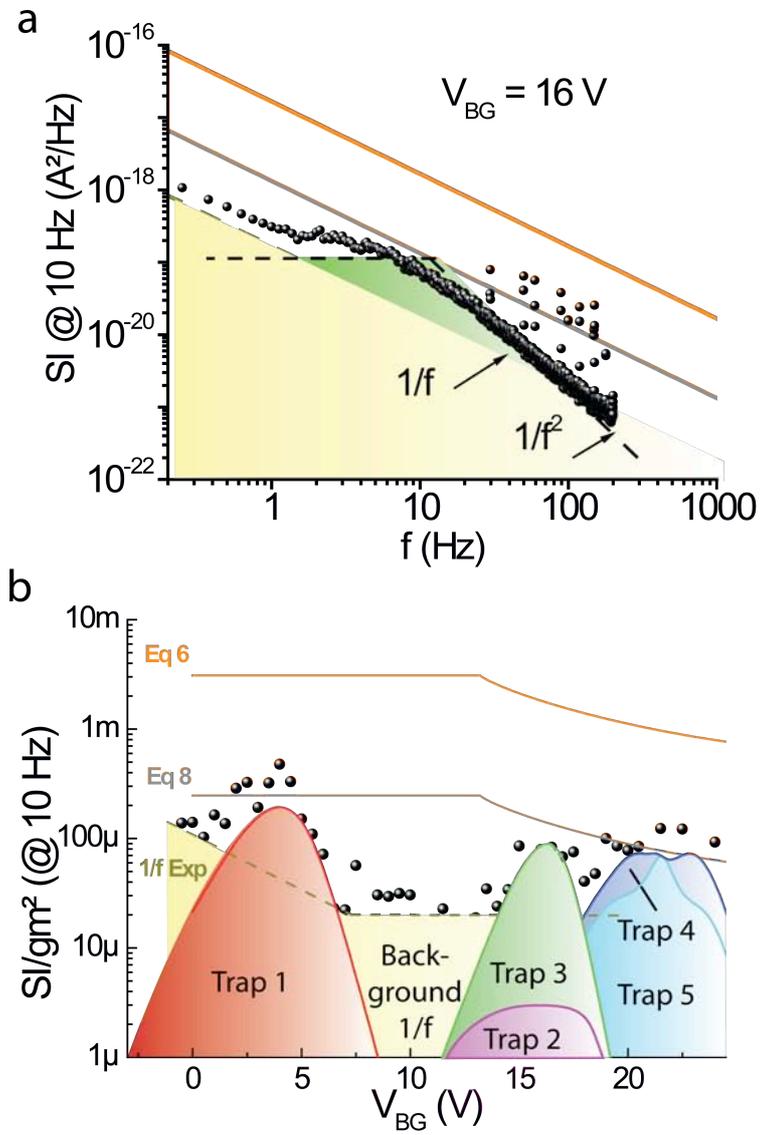

Fig.5

# One-by-one trap activation in silicon nanowire transistors

N. Clément[1], K.Nishiguchi[2], A.Fujiwara[2], & D. Vuillaume[1]

*(1) Institute of Electronics, Microelectronics and Nanotechnology, CNRS, Avenue Poincaré, 59652, Villeneuve d'Ascq, France*

*(2) NTT Basic Research Laboratories, 3-1, Morinosato Wakamiyia, Atsugi-shi, 243-0198 Japan*

## Supplementary Figures

**Supplementary Figure S1 Traps 4 and 5 occupancy probabilities using Eq.2**

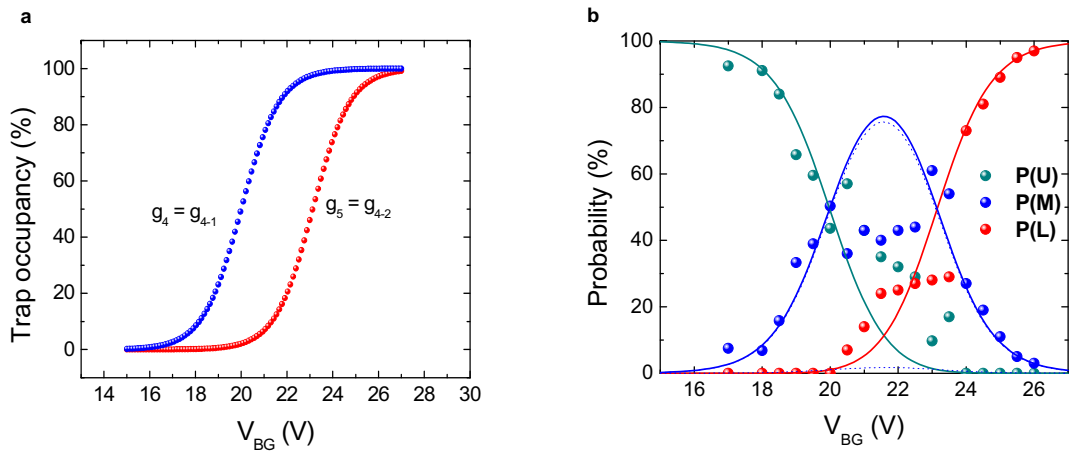

(a) *Trap occupancy considering asymptotes $g_4$ (blue) and $g_5$ (red)* (b) *Corresponding probability of being in the upper, middle and lower state. Experimental P(U), P(M) and P(L) (closed circles) are in disagreement with the theoretical curves in the three levels RTS $V_{BG}$ range (20..23.5 V)*

# Supplementary Figure S2: Other si nanowire showing one-by-one trap activation

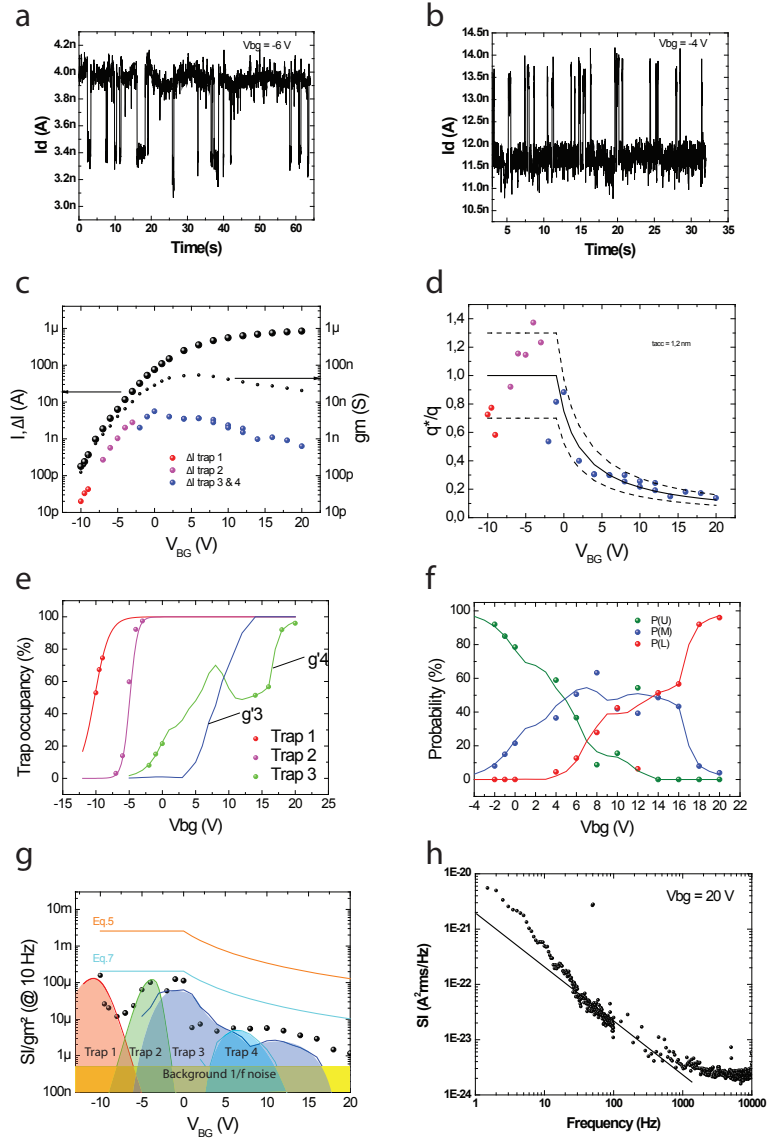

*Examples of RTSs at -6 and -4 V are shown in (a) and (b) (T=293 K). (c) shows drain current Id, RTS amplitude ΔI for 4 traps and transconductance. As for the other device, the flat band voltage fluctuation ΔVfb = ΔI/gm ≈ cte from the subthreshold to saturation operating regime. Using ΔVfb = q\*/Cg with q\*=q at V<Vth the threshold voltage, we obtain Cg = 0.72 aF and plot q\*/q in (d). Using eq. 5 with tacc = 1.2 nm we obtain a good fit for q\*/q (dashed curves are fits +/- 20 %). Trap occupancy probabilities for four traps and the probability of being in the upper, middle and lower state for three-level RTSs are plotted in (e) and (f), with the same procedure described in the main article.*

*Finally, an example of Lorentzian spectra at $V_{BG}$ = -1 V is shown in (h) and experimental input gate voltage referred noise SI/gm² (g) is fitted with the contribution of each trap separately (Eq.7). Simulation with Eq. 6 and 8 are also plotted as comparison. It confirms that eq.8 is a good approximation as an upper limit of noise.*

## Supplementary Figure S3: A silicon nanowire transistor with a single trap

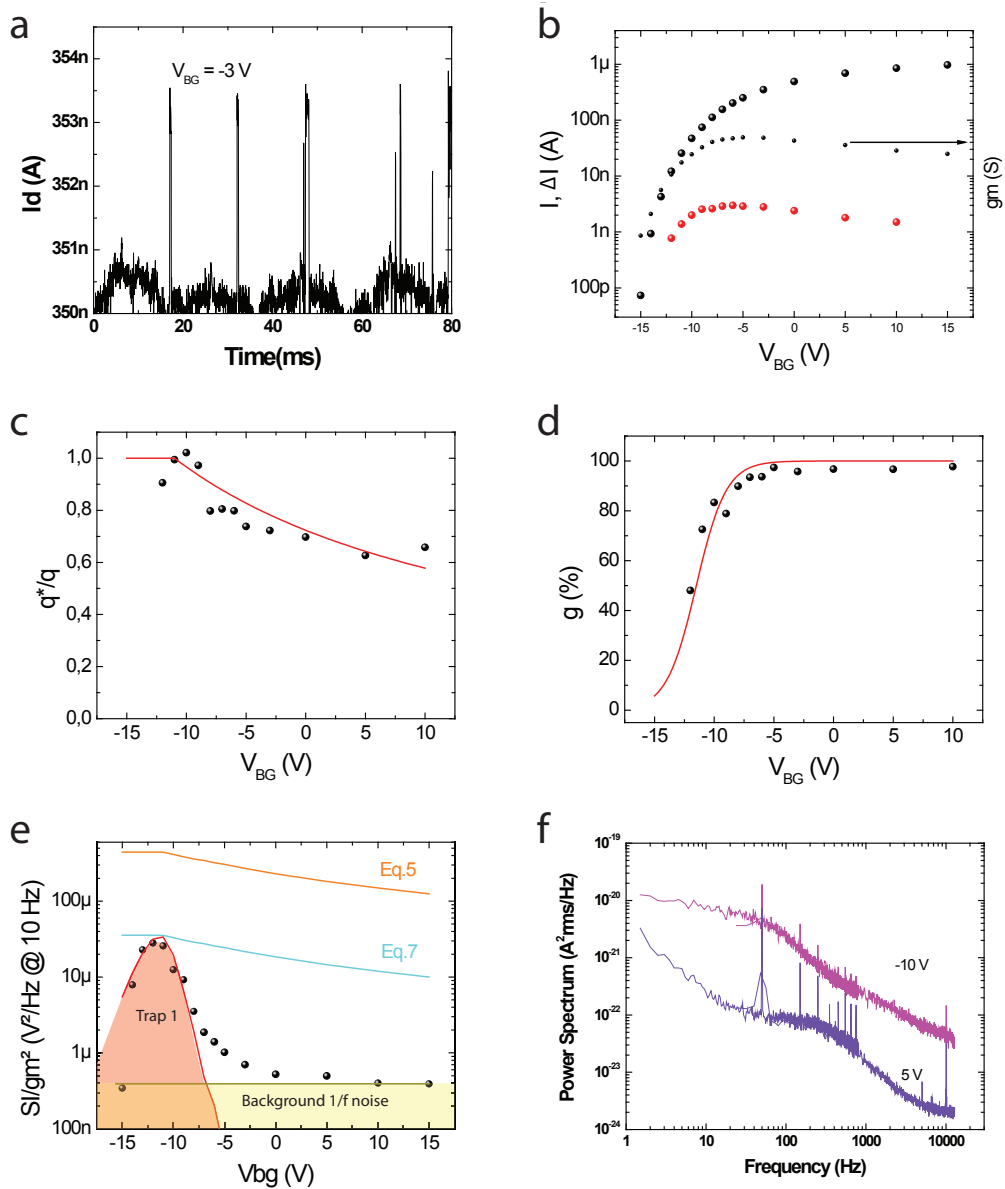

(a) *Example of RTSs at -3 and -4 V is shown in a (T=293 K). (b) shows drain current Id, RTS amplitude ΔI for the trap and transconductance. As for the other device, the flat band voltage fluctuation ΔVfb = ΔI/gm ≈ cte from the subthreshold to saturation operating regime. Using ΔVfb = q\*/Cg with q\*=q at V<Vth the threshold voltage, we obtain Cg = 2 aF and plot q\*/q in (c). Using eq. 5 with tacc = 0.14 nm, we obtain a good fit is obtained for q\*/q (red curve is fit). Trap occupancy probability is plotted in (d). In (e), experimental input gate voltage referred noise SI/gm² is fitted with the contribution of the unique trap (eq.7). Simulation with eqs.6 and 8 are also plotted for comparison. The results confirm that eq.8 is a good approximation as an upper limit of noise. An example of Lorentzian spectrum at $V_{BG}$ = -1 V is shown in (f).*